\documentclass[12pt,a4paper,final]{iopart}
\usepackage{iopams}  
\usepackage{graphicx}
\usepackage{multirow}
\usepackage{epsfig}
\usepackage{figsize}
\usepackage[breaklinks=true,colorlinks=true,linkcolor=blue,urlcolor=blue,citecolor=blue]{hyperref}
\usepackage{cite}

\begin{document}

\title[Spin bond order driven by extended repulsive interactions in doped graphene]
{Spin bond order driven by extended repulsive interactions in doped graphene}

\author{Jin-Ju Ri$^1$, Song-Jin O$^{2,3,*}$, Chol-Su O$^{1,3}$}

\address{$^1$Faculty of Energy Science, Kim Il Sung University, Taesong District, Pyongyang, Democratic People's Republic of Korea}
\address{$^2$Faculty of Physics, Kim Il Sung University, Taesong District, Pyongyang, Democratic People's Republic of Korea}
\address{$^3$Natural Science Center, Kim Il Sung University, Taesong District, Pyongyang, Democratic People's Republic of Korea}
\ead{sj.o@ryongnamsan.edu.kp}

\date{\today}


\begin{abstract}
	
We use the truncated-unity functional renormalization group (TUFRG) to study many-body instabilities of correlated electrons in graphene doped near the van Hove singularity (VHS). The system is described by an extended Hubbard model including several Coulomb repulsions between neighboring sites. With the repulsion parameters, which turn out to be suitable for low-energy consideration of graphene, we find a spin-bond-order phase in the vicinity of the VHS. This phase gives way to a spin-density-wave phase when involving a weak additional screening. The ground-state phase diagram is presented in the space of the doping level and the screening parameter. We describe in detail both of these spin-ordered states by using recently developed TUFRG + MF scheme, i.e., a combined approach of TUFRG and mean-field (MF) theory. The collinear states are energetically preferable in both cases of the spin bond order and the spin-density wave. But if the third-nearest-neighbor hopping is absent, these spin orders become chiral. The band structures of two collinear spin-ordered states are presented, revealing the metallic behavior of the system.

\end{abstract}

\vspace{2pc}

\submitto{\JPCM}

\maketitle

\section{Introduction}

After the experimental realization of graphene\cite{ref01}, this two-dimensional (2D) material has attracted considerable experimental and theoretical interest. In graphene, a conduction band of the $p_z$ electrons touches a valence band at the two $\bf{K}$ points. Therefore, many electronic properties of graphene near half-filling can be explained very well by a description in terms of 2D Dirac electrons\cite{ref02}. Such massless Dirac electrons exhibit some fascinating properties including the plasmonic Doppler effect\cite{ref02a}, the large diamagnetism originated from the electromagnetic duality\cite{ref02b}, the field-induced\cite{ref02e} and the spontaneous\cite{ref02c} valley polarizations, and the valley-related multiple topological phase transitions\cite{ref02d}. Involving weakly screened Coulomb interaction into the 2D system of the Dirac electrons leads to anomalous thermodynamic and magnetic properties\cite{ref03, ref04, ref05}.

Furthermore, doping graphene is an effective way for tailoring the electronic properties of this material. Doped graphene has been proposed as a candidate for some technological applications, such as an electrode of lithium-ion batteries\cite{ref06} and a photodetector\cite{ref07}. It can serve as a good platform for exploring various phase transitions. In general, when a system is doped close to the van Hove singularity (VHS), the effect of interactions could be strongly enhanced due to the divergent density of states at the Fermi level\cite{ref08}. Therefore, the heavily doped graphene near the VHS filling can host exotic ordered states\cite{ref09}.

In previous studies on graphene doped close to the VHS, i.e., near quarter doping, various electronic orders were identified using diverse approaches. The most remarkable result is the possibility of the electron-driven chiral $d$-wave superconductivity (SC), predicted in the random phase approximation\cite{ref10}, the Grassmann tensor product state approach\cite{ref11}, the finite-temperature determinantal quantum Monte Carlo (QMC)\cite{ref12, ref13}, and several renormalization group (RG) studies\cite{ref14, ref15, ref16}.

Near the VHS filling, the Fermi surface of graphene is nearly nested, which implies that spin-ordered phases can emerge, competing with the SC instability. Typically, two kinds of spin-density-wave (SDW) states have been proposed in previous works. For quarter-doped graphene, a chiral SDW state with nontrivial topology has been found by the mean-field (MF) theory\cite{ref17}, the singular-mode functional renormalization group (FRG)\cite{ref16}, and the finite-temperature determinantal QMC\cite{ref12}. Another mean-field theory\cite{ref18} has reported a collinear SDW state.

Most of these works have addressed the problem in a rather qualitative way. The results are based on simplified model Hamiltonians or biased considerations of special instability channels. In particular, the FRG studies were conducted without a quantitative description of the resulting symmetry-broken states. In this regard, a recently developed TUFRG + MF approach\cite{ref19} would be helpful. It is a combined approach of the truncated-unity FRG (TUFRG)\cite{ref20} and the MF theory, and has been applied to describe the half-filled\cite{ref21} and doped\cite{ref19} honeycomb lattices. In this work we employ the realistic Hamiltonian for graphene\cite{ref22} and map out the ground-state phase diagram. We focus on the spin-ordered states of quarter-doped graphene, presenting a detailed description of those states.

The rest of this paper is organized as follows: In Sec.~\ref{sec2}, we introduce the model for doped graphene, outline the TUFRG method, and present the phase diagram in the space of the doping level and the screening parameter. We then briefly explain in Sec.~\ref{sec3} the essence of the TUFRG + MF approach. In Sec.~\ref{sec4}, we describe in detail two kinds of emerging spin-ordered states, namely, the collinear spin-bond-order (SBO) and SDW phases. Finally, we summarize our results in Sec.~\ref{sec5}.

\section{Model, method and phase diagram} \label{sec2}

Doped graphene can be described by an extended Hubbard model of interacting spin-$1/2$ electrons with repulsions up to the third-nearest neighbor,
\begin{equation}\label{eq01}
H = H_0  + H_{{\rm int}} .
\end{equation}
Here the single-particle part $H_0$ includes hopping terms between the nearest ($t$), the second-nearest ($t_2$), and the third-nearest ($t_3$) neighbors,
\begin{eqnarray}\label{eq02}
\eqalign{
H_0  = & - t\sum\limits_{\left\langle {iA,jB} \right\rangle ,\sigma } {(c_{iA\sigma }^\dag  c_{jB\sigma }  + {\rm{H}}{\rm{.c}}{\rm{.}})}  - t_2 \sum\limits_{o,\left\langle {\left\langle {io,jo} \right\rangle } \right\rangle ,\sigma } {(c_{io\sigma }^\dag  c_{jo\sigma } }  + {\rm{H}}{\rm{.c}}{\rm{.}})\\
&- t_3 \sum\limits_{\left\langle {\left\langle {\left\langle {iA,jB} \right\rangle } \right\rangle } \right\rangle ,\sigma } {(c_{iA\sigma }^\dag  c_{jB\sigma } }  + {\rm{H}}{\rm{.c}}{\rm{.}}) - \mu \sum\limits_{i,o,\sigma } {c_{io\sigma }^\dag  c_{io\sigma } },
}
\end{eqnarray}
with the lattice site $i$ (or $j$), the sublattice index $o=A$ (or $B$), and the spin polarity $\sigma$. The interaction part $H_{{\rm int}}$ consists of the on-site repulsion ($U$) and Coulomb repulsive interactions between the nearest ($V_1$), the second-nearest ($V_2$), and the third-nearest ($V_3$) neighbors,
\begin{eqnarray}\label{eq03}
\eqalign{
H_{{\mathop{\rm int}} }  = & U\sum\limits_{i,o} {n_{io \uparrow } n_{io \downarrow } }  + V_1 \sum\limits_{\left\langle {iA,jB} \right\rangle } {\sum\limits_{\sigma ,\sigma '} {n_{iA\sigma } n_{jB\sigma '} } } \\
& + V_2 \sum\limits_{o,\left\langle {\left\langle {io,jo} \right\rangle } \right\rangle } {\sum\limits_{\sigma ,\sigma '} {n_{io\sigma } n_{jo\sigma '} } }  + V_3 \sum\limits_{\left\langle {\left\langle {\left\langle {iA,jB} \right\rangle } \right\rangle } \right\rangle } {\sum\limits_{\sigma ,\sigma '} {n_{iA\sigma } n_{jB\sigma '} } },
}
\end{eqnarray}
where $n_{io\sigma }  \equiv c_{io\sigma }^\dag  c_{io\sigma }$ is the local electron density operator for spin polarity $\sigma$.

From the experimental aspect, the unambiguous evidence of doping beyond the VHS in graphene was presented recently by employing a combination of ytterbium intercalation and potassium adsorption\cite{refA1}. In the case of such a chemical doping, the dopants can form a superlattice on top of graphene with a certain extent of randomness in their distribution, which would induce a periodic potential, change the lattice symmetry, and introduce a disorder, leading to a dopant-dependent strong renormalization of the bands\cite{refA2}. All these factors complicate an accurate prediction of the electronic structure. However, the electrostatic doping\cite{ref23} can overcome these drawbacks. Although, to our knowledge, the VHS filling of graphene has not yet been achieved in the electrostatic doping technique, we believe that it would be realized in future without a significant modification of the electronic bands.

Based on this argument, we assume in this work that the band structure of graphene would not be sizably renormalized when the electrostatic doping technique\cite{ref23} is used. Varying the gate voltage is equivalent to modifying the chemical potential $\mu$, which in turn change the number of electrons per atom $n_e$, controlling the doping level $\delta  \equiv n_e-1$. Following Refs. \cite{ref02, ref16}, we take $t = 2.8{\rm{eV}},t_2  = 0.1{\rm{eV}}$, and $t_3  = 0.07{\rm{eV}}$ for the hopping parameters. In this case, the parameters, $n_e$ and $\mu$, take the values of $n_e  \approx 1.25$ and $\mu _{{\rm{VHS}}}  = t + 2t_2  - 3t_3$ at the VHS filling.

For the repulsion parameters, we take $U = 9.3{\rm{eV}},V_1  = 5.5{\rm{eV}},V_2  = 4.1{\rm{eV}}$, and $V_3  = 3.6{\rm{eV}}$, as suggested in Ref. \cite{ref22}. Moreover, we take into account the additional screening due to the high-energy $\sigma$ bands and the electrons on the gate electrode. This effect can be parametrized by a substitution of repulsive interaction, $V({\bf{r}}) \to V({\bf{r}})e^{ - \gamma |{\bf{r}}|}$. Here we regard the exponent $\gamma$ as a phenomenological parameter. Thus we should modify the repulsion parameters as
\begin{eqnarray}\label{eq04}
\eqalign{
& U \to U,V_1  \to V_1 e^{ - \gamma r_1 } ,V_2  \to V_2 e^{ - \gamma r_2 }  = V_2 e^{ - \sqrt 3 \gamma r_1 } ,\\
& V_3  \to V_3 e^{ - \gamma r_3 }  = V_3 e^{ - 2\gamma r_1 } .
}
\end{eqnarray}
Introducing a screening parameter $\alpha  \equiv e^{ - \gamma r_1 }$, we can represent the renormalized repulsion parameters by
\begin{eqnarray}\label{eq05}
\eqalign{
\tilde V_1  = V_1 \alpha ,\tilde V_2  = V_2 \alpha ^{\sqrt 3 } ,\tilde V_3  = V_3 \alpha ^2 .
}
\end{eqnarray}
By replacing $V_i$ with $\tilde V_i$ in Eq. (\ref{eq03}), one can construct the interaction Hamiltonian for doped graphene, and use the TUFRG to analyze the ordering tendencies of the system.

As a flexible and highly scalable numerical scheme, the TUFRG\cite{ref20} is a recent version of the diagrammatically unbiased FRG method\cite{ref24, ref25, ref26}. It combines an efficient representation of the effective interaction in the exchange parametrization FRG\cite{ref27} and a simplified structure of the flow equation in the singular-mode FRG\cite{ref15}. This approach has been applied to analyze the ordering tendencies in various systems of interacting electrons\cite{ref28, ref29, ref30, ref31, ref32, ref33, ref34, ref35, ref36, ref37, ref38, ref39, ref40, ref41, ref42, ref43}.

In the FRG study, one traces an evolution of one-particle irreducible vertices, by introducing a scale-dependent cutoff ($G_0^\Omega  (\omega ,{\bf{k}})$) in the single-particle propagator. A level-2 truncated formalism of FRG is generally used. In this truncation one neglects all three and more particle vertices. Furthermore, we discard self-energy correction and frequency dependence of two-particle vertex (effective interaction) in this study. Thus, we consider the FRG flow equation of the effective interaction $V_{o_1 o_2 ,o_3 o_4 }^\Omega  ({\bf{k}}_1 ,{\bf{k}}_2 ;{\bf{k}}_3 ,{\bf{k}}_4 )$ to investigate spin-SU(2)-invariant systems. The interaction $V^\Omega$ consists of the initial interaction $V^{(0)}$ and three contributions from the particle-particle ($\Phi ^{{\rm{pp}}} (\Omega)$), crossed particle-hole ($\Phi ^{{\rm{ph,cr}}} (\Omega )$), and direct particle-hole ($\Phi ^{{\rm{ph,d}}} (\Omega )$) channels.
\begin{eqnarray}\label{eq06}
\eqalign{
V^\Omega   = V^{(0)}  + \Phi ^{{\rm{pp}}} (\Omega ) + \Phi ^{{\rm{ph,cr}}} (\Omega ) + \Phi ^{{\rm{ph,d}}} (\Omega )
}
\end{eqnarray}
In the TUFRG, each contributions are represented in terms of three matrices (bosonic propagators), $P^\Omega, C^\Omega$, and $D^\Omega$.
\begin{eqnarray}\label{eq07}
\eqalign{
\Phi ^{{\rm{pp}}} (\Omega ) \approx {\rm{\hat P}}^{ - 1} [P^\Omega  ],\Phi ^{{\rm{ph,cr}}} (\Omega ) \approx {\rm{\hat C}}^{ - 1} [C^\Omega  ],\Phi ^{{\rm{ph,d}}} (\Omega ) \approx {\rm{\hat D}}^{ - 1} [D^\Omega  ].
}
\end{eqnarray}
One can also express the effective interaction in a similar way,
\begin{eqnarray}\label{eq08}
\eqalign{
V^\Omega   \approx {\rm{\hat P}}^{ - 1} [V^{\rm{P}} (\Omega )] \approx {\rm{\hat C}}^{ - 1} [V^{\rm{C}} (\Omega )] \approx {\rm{\hat D}}^{ - 1} [V^{\rm{D}} (\Omega )].
}
\end{eqnarray}
The detailed expressions of Eqs. (\ref{eq07}) and (\ref{eq08}) are given in Ref. \cite{ref32}. By combining Eqs. (\ref{eq06}) and (\ref{eq07}) with the inverse of Eq. (\ref{eq08}), one can represent $V^{\rm{X}} (\Omega )$ $({\rm{X}} \in {\rm{P}},{\rm{C}},{\rm{D}})$ by the bosonoc propagators\cite{ref31, ref32}.

Then, the FRG flow equation of the effective interaction is transformed into the TUFRG flow equation for three bosonic propagators\cite{ref20, ref19}:
\begin{eqnarray}\label{eq09}
\eqalign{
& \frac{{dP^\Omega  ({\bf{q}})}}{{d\Omega }} = V^{{\rm{P}}(\Omega )} ({\bf{q}})\left[ {\frac{d}{{d\Omega }}\chi ^{{\rm{pp}}(\Omega )} ({\bf{q}})} \right]V^{{\rm{P}}(\Omega )} ({\bf{q}}),\\
& \frac{{dC^\Omega  ({\bf{q}})}}{{d\Omega }} = V^{{\rm{C}}(\Omega )} ({\bf{q}})\left[ {\frac{d}{{d\Omega }}\chi ^{{\rm{ph}}(\Omega )} ({\bf{q}})} \right]V^{{\rm{C}}(\Omega )} ({\bf{q}}), \\
& \frac{{d\left[ {C^\Omega  ({\bf{q}}) - 2D^\Omega  ({\bf{q}})} \right]}}{{d\Omega }} = \left[ {V^{{\rm{C}}(\Omega )} ({\bf{q}}) - 2V^{{\rm{D}}(\Omega )} ({\bf{q}})} \right] \\
& \hspace{5pc} \times \left[ {\frac{d}{{d\Omega }}\chi ^{{\rm{ph}}(\Omega )} ({\bf{q}})} \right]\left[ {V^{{\rm{C}}(\Omega )} ({\bf{q}}) - 2V^{{\rm{D}}(\Omega )} ({\bf{q}})} \right].
}
\end{eqnarray}
Here $\chi ^{{\rm{pp}}} (\Omega )$ and $\chi ^{{\rm{ph}}} (\Omega )$ are the particle-particle and particle-hole susceptibility matrices, respectively, and their definitions are presented in Ref. \cite{ref19}. In numeric implementation, we integrate this equation until we encounter a divergence in one of the three bosonic propagators. The divergence is associated with a phase transition that develops an order in the corresponding channel. Information on the type of order can be extracted from the singular modes of the divergent propagator.

\begin{figure}[h!]
	\begin{center}
		\includegraphics[width=8.6cm]{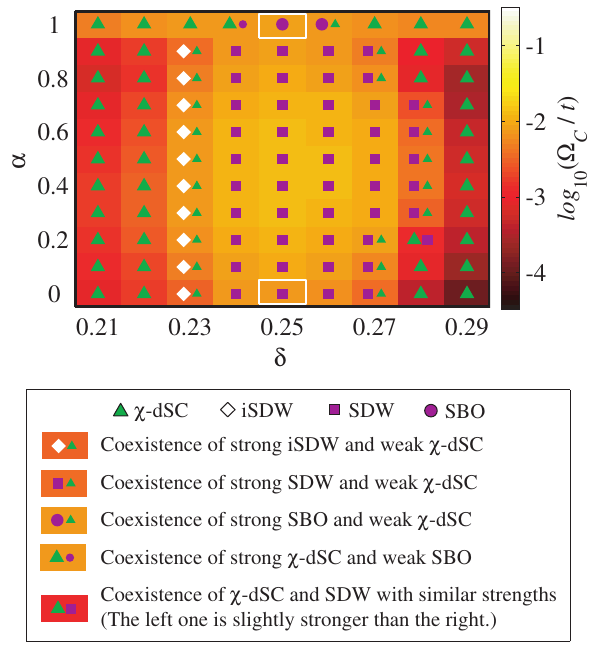}
	\end{center}
	\caption{Schematic ground-state phase diagram in the space of parameters $\alpha$ and $\delta$. As the estimates for the transition temperatures, the critical scales $\Omega_C$ are indicated using the color bar. The coexistence phases were identified via the same criterion as in Ref. \cite{ref32}.}
	\label{fig1}
\end{figure}

In our calculation, we only consider the bosonic propagators within the irreducible region (a triangle surrounded by a border line ${\bf{\Gamma }} - {\bf{K}} - {\bf{M}}$) of the Brillouin zone (BZ). The propagators outside this region can be generated using the point-group symmetry relations\cite{ref31} and the filtering process\cite{ref32}, which reduces the numerical effort by a factor of twelve. We have scanned the region of the parameter space, $\alpha  = 0.0$--$1.0$ and $\delta  = 0.21$--$0.29$, which includes the VHS filling $\delta _{{\rm{VHS}}}  \approx 0.25$. Through the TUFRG calculations for each parameter sets, we identify the ordering types, and also determine the critical scales $\Omega _C$ at which some divergence of the bosonic propagators is detected. The results are summarized in the schematic phase diagram of Fig. \ref{fig1}. The critical scales $\Omega _C$, which can be interpreted as an estimate for the transition temperature, are also presented using the color bar. The phase diagram contains the two-fold degenerate $d$-wave SC ($\chi$-dSC), the SDW, the SBO, and the incommensurate SDW (iSDW) phases. The two spin-ordered phases, the SDW and the SBO, are found at the transfer momentum $\bf{M}$.

In the absence of the additional screening ($\alpha  = 1$) we encounter an anomalous SBO state at quarter doping, which has not been reported before. However, it quickly turns into the SDW when involving a weak screening. In the case of localized repulsion ($\alpha=0$), the SDW and the $d$-wave SC become main ingredients of the phase diagram, with the incommensurate SDW emerging near $\delta  \approx 0.23$. This result is consistent with that of Ref. \cite{ref32}. The detailed form of linear combinations of degenerate singular modes can be determined only when addressing the symmetry-broken states. In the present study, this task was accomplished by using a recently developed TUFRG + MF approach\cite{ref19}, which is discussed in the following section. Our analysis shows that the two-fold degenerate SC modes constitute the chiral $d$-wave SC (denoted as $\chi$-dSC in Fig. \ref{fig1}), while for the SDW and the SBO, three degenerate spin modes at three partners of transfer momenta, ${\bf{M}}_1 ,{\bf{M}}_2$, and ${\bf{M}}_3$, build the collinear spin orders, not the chiral ones. Quantitative considerations of two representative states of the SDW and SBO phases, which correspond to two segments surrounded by white rectangular borders in Fig. \ref{fig1}, are presented in Sec.~\ref{sec4}.

\section{TUFRG + MF approach}\label{sec3}

The novel TUFRG + MF scheme\cite{ref19}, proposed by one of us, is an extension of the efficient FRG + MF approach\cite{ref44} to the TUFRG. In this approach, only the channel-irreducible part of the two-particle vertex resulting from the FRG calculation is taken as an input for the MF treatment. In the TUFRG + MF, bosonic propagators of the effective interaction are evolved by the TUFRG flow at high scale $\Omega  \ge \Omega _C$, but they change according to the random phase approximation (RPA) in the divergent regime ($\Omega  < \Omega _C$). Therefore, the propagators at low scale $\Omega  < \Omega _C$ will be identical to those obtained by using only the RPA starting from the irreducible bosonic propagators, $\tilde P,\tilde C$, and $\tilde D$.

Taking into account the equivalence of two critical conditions in the RPA and the MF theory\cite{ref19}, the interaction Hamiltonian corresponding to the irreducible bosonic propagators is used as an input interaction for the MF calculation. Since the FRG flow equation is replaced by the matrix flow equation in the TUFRG, the relations between the bosonoc and the irreducible bosonic propagators are expressed by concise matrix equations, which read,
\begin{eqnarray}\label{eq10}
\eqalign{
[ - \tilde P({\bf{q}})]^{ - 1} & = [ - P^{\Omega _C } ({\bf{q}})]^{ - 1}  - \chi ^{{\rm{pp}}(\Omega _C )} ({\bf{q}}),\\
[\tilde C({\bf{q}})]^{ - 1} & = [C^{\Omega _C } ({\bf{q}})]^{ - 1} + \chi ^{{\rm{ph}}(\Omega _C )} ({\bf{q}}),\\
[\tilde W({\bf{q}})]^{ - 1} & = [W^{\Omega _C } ({\bf{q}})]^{ - 1}  + \chi ^{{\rm{ph}}(\Omega _C )} ({\bf{q}}),
}
\end{eqnarray}
with a definition
\begin{eqnarray}\label{eq11}
\eqalign{
\tilde W({\bf{q}}) \equiv \tilde C({\bf{q}}) - 2\tilde D({\bf{q}}),W^{\Omega _C } ({\bf{q}}) \equiv C^{\Omega _C } ({\bf{q}}) - 2D^{\Omega _C } ({\bf{q}}).
}
\end{eqnarray}
At the critical scale, some bosonic propagators exhibit strong divergence at particular transfer momenta ${\bf{Q}}_i$, and they can be approximately expanded in terms of several singular modes associated with dominant positive eigenvalues.
\begin{eqnarray}\label{eq12}
\eqalign{
- P^{\Omega _C } ({\bf{Q}}_i^{\rm{P}} ) &= \sum\limits_{\alpha  = 1}^{M_{{\rm{P}},i} } {\lambda ^{{\rm{P}},\alpha } ({\bf{Q}}_i^{\rm{P}} )} \left| {\phi ^{{\rm{P}},\alpha } ({\bf{Q}}_i^{\rm{P}} )} \right\rangle \left\langle {\phi ^{{\rm{P}},\alpha } ({\bf{Q}}_i^{\rm{P}} )} \right|,\\
C^{\Omega _C } ({\bf{Q}}_i^{\rm{C}} ) &= \sum\limits_{\alpha  = 1}^{M_{{\rm{C}},i} } {\lambda ^{{\rm{C}},\alpha } ({\bf{Q}}_i^{\rm{C}} )} \left| {\phi ^{{\rm{C}},\alpha } ({\bf{Q}}_i^{\rm{C}} )} \right\rangle \left\langle {\phi ^{{\rm{C}},\alpha } ({\bf{Q}}_i^{\rm{C}} )} \right|,\\
W^{\Omega _C } ({\bf{Q}}_i^{\rm{W}} ) &= \sum\limits_{\alpha  = 1}^{M_{{\rm{W}},i} } {\lambda ^{{\rm{W}},\alpha } ({\bf{Q}}_i^{\rm{W}} )} \left| {\phi ^{{\rm{W}},\alpha } ({\bf{Q}}_i^{\rm{W}} )} \right\rangle \left\langle {\phi ^{{\rm{W}},\alpha } ({\bf{Q}}_i^{\rm{W}} )} \right|.
}
\end{eqnarray}
Here, e.g., $M_{{\rm{P}},i}$ is the number of the singular modes $\left| {\phi ^{{\rm{P}},\alpha } ({\bf{Q}}_i^{\rm{P}} )} \right\rangle$.

Then, the solution of Eq. (\ref{eq10}) reads
\begin{eqnarray}\label{eq13}
\eqalign{
- \tilde P({\bf{Q}}_i^{\rm{P}} ) &= \sum\limits_{\alpha  = 1}^{M_{{\rm{P}},i} } {\Lambda ^{{\rm{P}},\alpha } ({\bf{Q}}_i^{\rm{P}} )} \left| {\varphi ^{{\rm{P}},\alpha } ({\bf{Q}}_i^{\rm{P}} )} \right\rangle \left\langle {\varphi ^{{\rm{P}},\alpha } ({\bf{Q}}_i^{\rm{P}} )} \right|,\\
\tilde C({\bf{Q}}_i^{\rm{C}} ) &= \sum\limits_{\alpha  = 1}^{M_{{\rm{C}},i} } {\Lambda ^{{\rm{C}},\alpha } ({\bf{Q}}_i^{\rm{C}} )} \left| {\varphi ^{{\rm{C}},\alpha } ({\bf{Q}}_i^{\rm{C}} )} \right\rangle \left\langle {\varphi ^{{\rm{C}},\alpha } ({\bf{Q}}_i^{\rm{C}} )} \right|,\\
\tilde W({\bf{Q}}_i^{\rm{W}} ) &= \sum\limits_{\alpha  = 1}^{M_{{\rm{W}},i} } {\Lambda ^{{\rm{W}},\alpha } ({\bf{Q}}_i^{\rm{W}} )} \left| {\varphi ^{{\rm{W}},\alpha } ({\bf{Q}}_i^{\rm{W}} )} \right\rangle \left\langle {\varphi ^{{\rm{W}},\alpha } ({\bf{Q}}_i^{\rm{W}} )} \right|.
}
\end{eqnarray}
Here the irreducible coupling constant $\Lambda ^{{\rm{X}},\alpha } ({\bf{Q}}_i^{\rm{X}} )$ $({\rm{X}} \in {\rm{P}},{\rm{C}},{\rm{W}})$ is the eigenvalue of the matrix $Y^{\rm{X}} ({\bf{Q}}_i^{\rm{X}} )$ which is defined by
\begin{eqnarray}\label{eq14}
\eqalign{
[Y^{\rm{P}} ({\bf{Q}}_i^{\rm{P}} )^{ - 1} ]_{\alpha \alpha '}  &\equiv \left( {\frac{1}{{\lambda ^{{\rm{P}},\alpha } ({\bf{Q}}_i^{\rm{P}} )}}\delta _{\alpha \alpha '}  + \left\langle {\phi ^{{\rm{P}},\alpha } ({\bf{Q}}_i^{\rm{P}} )} \right|[ - \chi ^{{\rm{pp}}(\Omega _{\rm{D}} )} ({\bf{Q}}_i^{\rm{P}} )]\left| {\phi ^{{\rm{P}},\alpha '} ({\bf{Q}}_i^{\rm{P}} )} \right\rangle } \right), \\
[Y^{\rm{C}} ({\bf{Q}}_i^{\rm{C}} )^{ - 1} ]_{\alpha \alpha '}  &\equiv \left( {\frac{1}{{\lambda ^{{\rm{C}},\alpha } ({\bf{Q}}_i^{\rm{C}} )}}\delta _{\alpha \alpha '}  + \left\langle {\phi ^{{\rm{C}},\alpha } ({\bf{Q}}_i^{\rm{C}} )} \right|\chi ^{{\rm{ph}}(\Omega _{\rm{D}} )} ({\bf{Q}}_i^{\rm{C}} )\left| {\phi ^{{\rm{C}},\alpha '} ({\bf{Q}}_i^{\rm{C}} )} \right\rangle } \right), \\
[Y^{\rm{W}} ({\bf{Q}}_i^{\rm{W}} )^{ - 1} ]_{\alpha \alpha '}  &\equiv \left( {\frac{1}{{\lambda ^{{\rm{W}},\alpha } ({\bf{Q}}_i^{\rm{W}} )}}\delta _{\alpha \alpha '}  + \left\langle {\phi ^{{\rm{W}},\alpha } ({\bf{Q}}_i^{\rm{W}} )} \right|\chi ^{{\rm{ph}}(\Omega _{\rm{D}} )} ({\bf{Q}}_i^{\rm{W}} )\left| {\phi ^{{\rm{W}},\alpha '} ({\bf{Q}}_i^{\rm{W}} )} \right\rangle } \right).
}
\end{eqnarray}
And the irreducible singular mode $\left| {\varphi ^{{\rm{P}},\alpha } ({\bf{Q}}_i^{\rm{P}} )} \right\rangle$ is defined as
\begin{eqnarray}\label{eq15}
\eqalign{
\left| {\varphi ^{{\rm{X}},\alpha } ({\bf{Q}}_i^{\rm{X}} )} \right\rangle  \equiv \sum\limits_{\beta  = 1}^{M_{{\rm{X}},i} } {S_\beta ^{{\rm{X}},\alpha } ({\bf{Q}}_i^{\rm{X}} )} \left| {\phi ^{{\rm{X}},\beta } ({\bf{Q}}_i^{\rm{X}} )} \right\rangle ,
}
\end{eqnarray}
with the eigenvector ${\bf{S}}^{{\rm{X}},\alpha } ({\bf{Q}}_i^{\rm{X}} ) = (S_1^{{\rm{X}},\alpha } ({\bf{Q}}_i^{\rm{X}} ),S_2^{{\rm{X}},\alpha } ({\bf{Q}}_i^{\rm{X}} ), \cdots ,S_{M_{{\rm{X}},i} }^{{\rm{X}},\alpha } ({\bf{Q}}_i^{\rm{X}} ))$ of the matrix $Y^{\rm{X}} ({\bf{Q}}_i^{\rm{X}} )$, associated with the eigenvalue $\Lambda ^{{\rm{X}},\alpha } ({\bf{Q}}_i^{\rm{X}} )$.

The input interaction is given by
\begin{eqnarray}\label{eq16}
\eqalign{
V^{{\rm{MF}}}  = {\rm{\hat P}}^{ - 1} [\tilde P] + {\rm{\hat C}}^{ - 1} [\tilde C] + {\rm{\hat D}}^{ - 1} [\tilde D],
}
\end{eqnarray}
which is inserted into the irreducible Hamiltonian,
\begin{eqnarray}\label{eq17}
\eqalign{
H^{{\rm{irred}}}  =& H_0  + \frac{1}{{2N}}\sum\limits_{o_1 ,o_2 ,o_3 ,o_4 } {\sum\limits_{{\bf{k}}_1 ,{\bf{k}}_2 ,{\bf{k}}_3 } {\sum\limits_{\sigma ,\sigma '} {V_{o_1 o_2 ,o_3 o_4 }^{{\rm{MF}}} ({\bf{k}}_1 ,{\bf{k}}_2 ;{\bf{k}}_3 ,{\bf{k}}_4 )} } }\\
& \times c_{{\bf{k}}_1 o_1 \sigma }^\dag  c_{{\bf{k}}_2 o_2 \sigma '}^\dag  c_{{\bf{k}}_4 o_4 \sigma '} c_{{\bf{k}}_3 o_3 \sigma },\\
& \textrm{ with a constraint of} \hspace{1pc} {\bf{k}}_4  = {\bf{k}}_1  + {\bf{k}}_2  - {\bf{k}}_3.
}
\end{eqnarray}
In the mean-field approximation, it is replaced by the following MF Hamiltonian:
\begin{eqnarray}\label{eq18}
\eqalign{
H_{{\rm{MF}}}  = H_0  + H_{{\rm{sSC}}}  + H_{{\rm{tSC}}}  + H_{{\rm{SPN}}}  + H_{{\rm{CHG}}} .
}
\end{eqnarray}
The detailed expressions of $H_{{\rm{sSC}}} ,H_{{\rm{tSC}}} ,H_{{\rm{SPN}}}$, and $H_{{\rm{CHG}}}$ can be found in Ref. \cite{ref19}. All of these quantities include the order parameters $\Delta _\alpha ^{\rm{X}}$ $({\rm{X}} \in {\rm{sSC}},{\rm{tSC}},{\rm{SPN}},{\rm{CHG}})$, the fermion bilinear operators $O_\alpha ^{\rm{X}}$, and the irreducible coupling constants $\Lambda ^{{\rm{X}},\alpha }$, while the operator $O_\alpha ^{\rm{X}}$ contains the irreducible singular mode $\left| {\varphi ^{{\rm{X}},\alpha } } \right\rangle$ in it. Since fermion quartic terms are absent in $H_{{\rm{MF}}}$, the problem is exactly solvable. The order parameters are determined by the self-consistency condition,
\begin{eqnarray}\label{eq19}
\eqalign{
\Delta _\alpha ^{\rm{X}} ({\bf{Q}}) = \frac{{\Lambda ^{{\rm{X}},\alpha } ({\bf{Q}})}}{{4N}}\left\langle {O_\alpha ^{\rm{X}} ({\bf{Q}})} \right\rangle _{{\rm{MF}}}.
}
\end{eqnarray}
At zero temperature, this condition is identical to the requirement to minimize the ground-state energy $E_{\rm{GS}}$ of the MF Hamiltonian with respect to $\Delta _\alpha ^{\rm{X}}$, while keeping $n_e$ fixed.

\section{Analysis of SBO and SDW phases}\label{sec4}

In this section we focus on two states, namely, the SBO and the SDW states, which are marked by white rectangles in Fig. \ref{fig1}. In the TUFRG flows for these parameter sets, only the bosonic propagator $C^\Omega$ shows a strong divergence at the transfer momentum $\bf{M}$ (more precisely at ${\bf{M}}_1  = \frac{{2\pi }}{{\sqrt 3 a}}(0,1)$ with the lattice constant $a$), having the nondegenerate singular mode. This means that $Y^{\rm{C}} ({\bf{M}}_1 )$ is a real number and $\left| {\varphi ^{\rm{C}} ({\bf{M}}_1 )} \right\rangle  = \left| {\phi ^{\rm{C}} ({\bf{M}}_1 )} \right\rangle $. Then Eqs. (\ref{eq13}) and (\ref{eq14}) become
\begin{eqnarray}\label{eq20}
\eqalign{
\tilde C({\bf{M}}_1 ) &= \Lambda ^{\rm{C}} ({\bf{M}}_1 )\left| {\varphi ^{\rm{C}} ({\bf{M}}_1 )} \right\rangle \left\langle {\varphi ^{\rm{C}} ({\bf{M}}_1 )} \right|,\\
Y^{\rm{C}} ({\bf{M}}_1 ) &= \Lambda ^{\rm{C}} ({\bf{M}}_1 )\\
&= \left( {\frac{1}{{\lambda ^{\rm{C}} ({\bf{M}}_1 )}} + \left\langle {\varphi ^{\rm{C}} ({\bf{M}}_1 )} \right|\chi ^{{\rm{ph}}(\Omega _{\rm{D}} )} ({\bf{M}}_1 )\left| {\varphi ^{\rm{C}} ({\bf{M}}_1 )} \right\rangle } \right)^{ - 1} .
}
\end{eqnarray}
The propagator $C^\Omega$ also exhibits the identical divergences at another two partners, ${\bf{M}}_2  = \frac{{2\pi }}{{\sqrt 3 a}}\left( { - \frac{{\sqrt 3 }}{2}, - \frac{1}{2}} \right)$ and ${\bf{M}}_3  = \frac{{2\pi }}{{\sqrt 3 a}}\left( {\frac{{\sqrt 3 }}{2}, - \frac{1}{2}} \right)$. The irreducible singular modes, $\left| {\varphi ^{\rm{C}} ({\bf{M}}_2 )} \right\rangle$ and $\left| {\varphi ^{\rm{C}} ({\bf{M}}_3 )} \right\rangle$ can be obtained by applying point-group symmetry relations\cite{ref31} to the mode $\left| {\varphi ^{\rm{C}} ({\bf{M}}_1 )} \right\rangle$.

\begin{figure}[h!]
	\begin{center}
		\includegraphics[width=15.6cm]{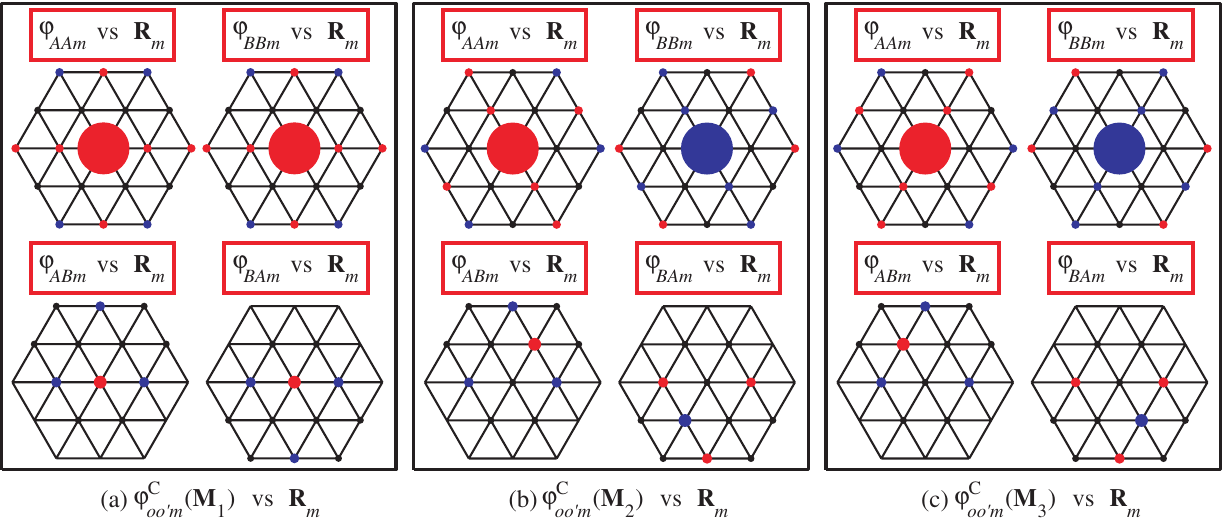}
	\end{center}
	\caption{Values of the irreducible singular modes in the spin channel, (a) $\varphi _{oo'm}^{\rm{C}} ({\bf{M}}_1)$, (b) $\varphi _{oo'm}^{\rm{C}} ({\bf{M}}_2)$, and (c) $\varphi _{oo'm}^{\rm{C}} ({\bf{M}}_3)$ for the SDW state ($\alpha  = 0.0,\delta  = 0.25$). The red and blue circles indicate the positive and negative values, respectively, and the absolute values $\left| {\varphi _{oo'm} } \right|$ are encoded by the radius of the circles. The small dots denote the sites ${\bf{R}}_m$ having negligible $\varphi _{oo'm}$, while the empty sites are eliminated by the filtering process\cite{ref32}.}
	\label{fig2}
\end{figure}

\begin{figure}[h!]
	\begin{center}
		\includegraphics[width=15.6cm]{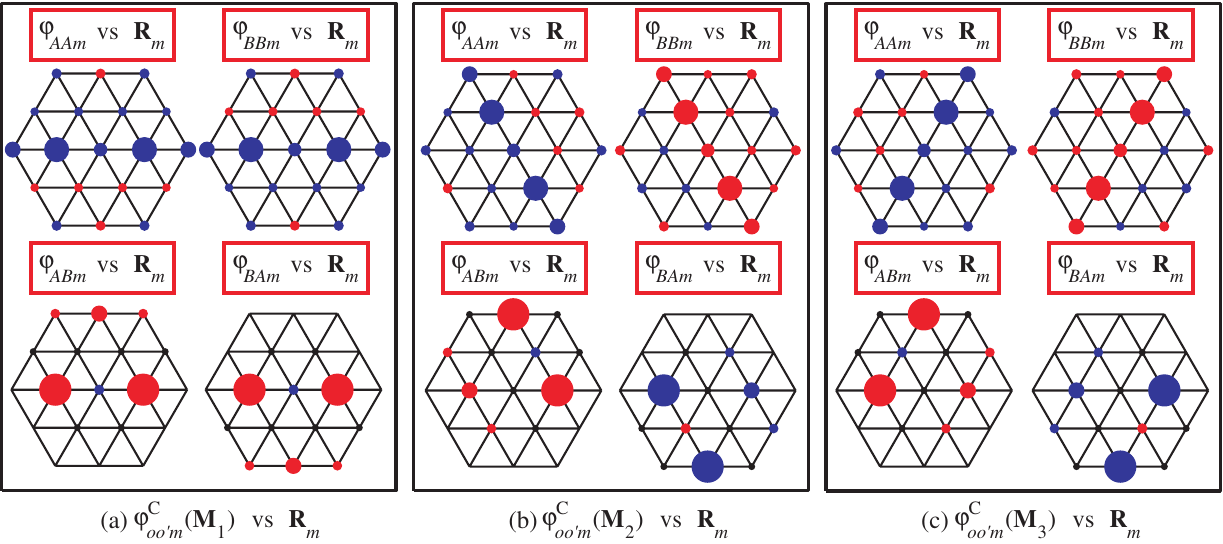}
	\end{center}
	\caption{Values of the irreducible singular modes in the spin channel, (a) $\varphi _{oo'm}^{\rm{C}} ({\bf{M}}_1)$, (b) $\varphi _{oo'm}^{\rm{C}} ({\bf{M}}_2)$, and (c) $\varphi _{oo'm}^{\rm{C}} ({\bf{M}}_3)$ for the SBO state ($\alpha  = 1.0,\delta  = 0.25$). The red and blue circles, the radius of them, the small dots, and the empty sites have the same meanings as in Fig. \ref{fig2}.}
	\label{fig3}
\end{figure}

Fig. \ref{fig2} shows three irreducible singular modes of the SDW state at $\alpha  = 0.0,\delta  = 0.25$, while Fig. \ref{fig3} shows three modes of the SBO state at $\alpha  = 1.0,\delta  = 0.25$. Note that all the elements of $\left| {\varphi ^{\rm{C}} ({\bf{M}}_i )} \right\rangle$ are real numbers. As will be demonstrated in later discussion, the strongest spin bonds are formed between the third-nearest neighbors in this SBO state.

The symmetry relations also give the following equalities:
\begin{eqnarray}\label{eq21}
\eqalign{
\lambda ^{\rm{C}} ({\bf{M}}_1 ) = \lambda ^{\rm{C}} ({\bf{M}}_2 ) = \lambda ^{\rm{C}} ({\bf{M}}_3 ) = \lambda ,\\
\Lambda ^{\rm{C}} ({\bf{M}}_1 ) = \Lambda ^{\rm{C}} ({\bf{M}}_2 ) = \Lambda ^{\rm{C}} ({\bf{M}}_3 ) = \Lambda ,\\
\left\langle {\varphi ^{\rm{C}} ({\bf{M}}_i )} \right|\chi ^{{\rm{ph}}(\Omega _{\rm{D}} )} ({\bf{M}}_i )\left| {\varphi ^{\rm{C}} ({\bf{M}}_i )} \right\rangle  = \chi  \hspace{1pc} (i = 1,2,3).	
}
\end{eqnarray}
The MF Hamiltonian in Eq. (\ref{eq18}) becomes $H_{{\rm{MF}}}  = H_0  + H_{{\rm{SPN}}}$, where the detailed expression of $H_{{\rm{SPN}}}$ reads (see Eqs. (86) and (87) of Ref. \cite{ref19}),
\begin{eqnarray}\label{eq22}
\eqalign{
H_{{\rm{SPN}}}  = \sum\limits_{i = 1}^3 {\left( {\frac{{4N}}{\Lambda }|\vec \Delta ^{{\rm{SPN}}} ({\bf{M}}_i )|^2  - 2\vec \Delta ^{{\rm{SPN}}} ({\bf{M}}_i ) \cdot \vec O^{{\rm{SPN}}} ({\bf{M}}_i )} \right)} ,\\
\vec O^{{\rm{SPN}}} ({\bf{M}}_i ) = \sum\limits_{\bf{k}} {\sum\limits_{o,o',m} {\left[ {\phi _{oo'm}^{\rm{C}} ({\bf{M}}_i )} \right]} } ^* e^{i{\bf{R}}_m  \cdot {\bf{k}}} \sum\limits_{\sigma ,\sigma '} {c_{{\bf{k}},o',\sigma }^\dag  \vec \sigma _{\sigma \sigma '} c_{{\bf{k}} + {\bf{M}}_i ,o,\sigma '} } .	
}
\end{eqnarray}
Here we used the relation,
\begin{eqnarray}\label{eq23}
\eqalign{
\vec \Delta ^{{\rm{SPN}}} ({\bf{M}}_i ) = [\vec \Delta ^{{\rm{SPN}}} ({\bf{M}}_i )]^* ,\vec O^{{\rm{SPN}}} ({\bf{M}}_i ) = [\vec O^{{\rm{SPN}}} ({\bf{M}}_i )]^\dag  ,
}
\end{eqnarray}
resulting from the physical equivalence of two vectors, ${\bf{M}}_i$ and $- {\bf{M}}_i$. The order parameters $\vec \Delta ^{{\rm{SPN}}} ({\bf{M}}_i )$ are determined by the minimization of the ground-state energy of $H_{{\rm{MF}}}$. Two candidates for them have been proposed, one is the chiral spin-ordered phase\cite{ref16, ref17} ($\vec \Delta ^{{\rm{SPN}}} ({\bf{M}}_1 ) = \Delta {\bf{e}}_z ,\vec \Delta ^{{\rm{SPN}}} ({\bf{M}}_2 ) = \Delta {\bf{e}}_x ,\vec \Delta ^{{\rm{SPN}}} ({\bf{M}}_3 ) = \Delta {\bf{e}}_y$), and another is the collinear phase\cite{ref18} ($\vec \Delta ^{{\rm{SPN}}} ({\bf{M}}_1 ) = \vec \Delta ^{{\rm{SPN}}} ({\bf{M}}_2 ) = \vec \Delta ^{{\rm{SPN}}} ({\bf{M}}_3 ) = \Delta {\bf{e}}_z$).

\begin{table}[h!]
	\caption{Results of the TUFRG + MF calculation.}
	\begin{center}
		\begin{tabular}{|c|c|c|c|c|c|c|}
			\hline
			Parameter & \multirow{2}{*} {$\lambda$ (eV)} & \multirow{2}{*} {$\chi$ (eV)} & \multirow{2}{*} {$\Lambda$ (eV)} & \multirow{2}{*} {Type of order} & \multirow{2}{*} {$\Delta$ (eV)} & $E_{{\rm{GS}}}(\Delta)-E_{{\rm{GS}}}(0)$ \\
			set & & & & & & (eV/Cell) \\  \hline
			$\alpha  = 0.0,$ & \multirow{2}{*} {380.161} & \multirow{2}{*} {0.15197} & \multirow{2}{*} {6.46842} & Collinear SDW & $6.417 \times 10^{-2}$ & $-4.853 \times 10^{-4}$ \\  \cline{5-7}
			$\delta  = 0.25.$ & & & & Chiral SDW & $6.003 \times 10^{-2}$ & $-4.340 \times 10^{-4}$ \\  \hline
			$\alpha  = 1.0,$ & \multirow{2}{*} {1298.27} & \multirow{2}{*} {0.36389} & \multirow{2}{*} {2.74229} & Collinear SBO & $4.212 \times 10^{-2}$ & $-7.840 \times 10^{-4}$ \\  \cline{5-7}
			$\delta  = 0.25.$ & & & & Chiral SBO & $4.124 \times 10^{-2}$ & $-6.910 \times 10^{-4}$ \\
			\hline
		\end{tabular}
	\end{center}
	\label{tab01}
\end{table}

\begin{table}[h!]
	\caption{Results of the TUFRG + MF calculation for vanishing $t_3$.}
	\begin{center}
		\begin{tabular}{|c|c|c|c|c|c|c|}
			\hline
			Parameter set & \multirow{2}{*} {$\lambda$ (eV)} & \multirow{2}{*} {$\chi$ (eV)} & \multirow{2}{*} {$\Lambda$ (eV)} & \multirow{2}{*} {Type of order} & \multirow{2}{*} {$\Delta$ (eV)} & $E_{{\rm{GS}}}(\Delta)-E_{{\rm{GS}}}(0)$ \\
			(with $t_3$ set to 0) & & & & & & (eV/Cell) \\  \hline
			$\alpha  = 0.0,$ & \multirow{2}{*} {406.569} & \multirow{2}{*} {0.15455} & \multirow{2}{*} {6.36911} & Collinear SDW & $6.522 \times 10^{-2}$ & $-1.411 \times 10^{-3}$ \\  \cline{5-7}
			$\delta  = 0.25.$ & & & & Chiral SDW & $7.133 \times 10^{-2}$ & $-1.557 \times 10^{-3}$ \\  \hline
			$\alpha  = 1.0,$ & \multirow{2}{*} {7767.67} & \multirow{2}{*} {0.39266} & \multirow{2}{*} {2.54589} & Collinear SBO & $3.748 \times 10^{-2}$ & $-1.518 \times 10^{-3}$ \\  \cline{5-7}
			$\delta  = 0.25.$ & & & & Chiral SBO & $3.992 \times 10^{-2}$ & $-1.653 \times 10^{-3}$ \\
			\hline
		\end{tabular}
	\end{center}
	\label{tab01A}
\end{table}

The calculation results are summarized in Table \ref{tab01}. As shown in the table, our results point out the collinear spin orders for both states, unlike the previous results\cite{ref12, ref16, ref17, ref19}. To reveal the effect of the hopping parameter $t_3$, we performed the TUFRG + MF calculation for both parameter sets with the hopping changed to $t_3=0$. The calculation results are summarized in Table \ref{tab01A}. As can be seen from this table, the chiral spin orders are favored for both cases, which is consistent with the previous results\cite{ref12, ref16, ref17, ref19}. Thus, if one eliminates the third-nearest-neighbor hopping, while keeping other parameters unchanged, then the resulting spin orders change from the collinear to the chiral ones. This means that the chiral and collinear orders are easily transformed into each other, depending on the shape of the Fermi surface. A detailed analysis of the effect of $t_3$ will be presented below.

The spin part of the MF Hamiltonian, $H_{{\rm{SPN}}}$ in Eq. (\ref{eq22}), reduces the BZ by a factor of four. The band structures of the collinear SDW and SBO states can be displayed in this small BZ. Since the spin projection along $z$-axis ($S_z$) is conserved in the collinear states, it serves as a good quantum number and the bands can be divided into two groups of spin-up and spin-down electrons. Spin-up electrons possess the spin projection along the direction ($+z$) of the order parameter $\vec \Delta ^{{\rm{SPN}}} ({\bf{M}}_i )$. The Fermi surface is shared by the 5-th and 6-th bands. The characteristics of the band structures are listed in Table \ref{tab02}.

\begin{table}[h!]
	\caption{Characteristics of the band structures for the SDW and the SBO states.}
	\begin{center}
		\begin{tabular}{|c|c|c|c|c|c|}
			\hline
			 \multirow{2}{*}{Parameter set} & \multicolumn{2}{|c|}{Top of 5-th band (eV)} & \multicolumn{2}{|c|}{Bottom of 6-th band (eV)} & $\mu$ in MF \\ \cline{2-5}
			& spin-up & spin-down & spin-up & spin-down & calc. (eV) \\
			\hline
			$\alpha  = 0.0, \delta  = 0.25.$ & \multirow{2}{*}{2.85387} & \multirow{2}{*}{2.85284} & \multirow{2}{*}{2.85859} & \multirow{2}{*}{2.72141} & \multirow{2}{*}{2.832} \\
			(Collinear SDW) &  &  &  &  & \\
			\hline
			$\alpha  = 1.0, \delta  = 0.25.$ & \multirow{2}{*}{2.83686} & \multirow{2}{*}{2.83704} & \multirow{2}{*}{2.86815} & \multirow{2}{*}{2.71185} & \multirow{2}{*}{2.817} \\
			(Collinear SBO) &  &  &  &  & \\
			\hline
		\end{tabular}
	\end{center}
	\label{tab02}
\end{table}
	
\begin{figure}[h!]
	\begin{center}
		\includegraphics[width=14.0cm]{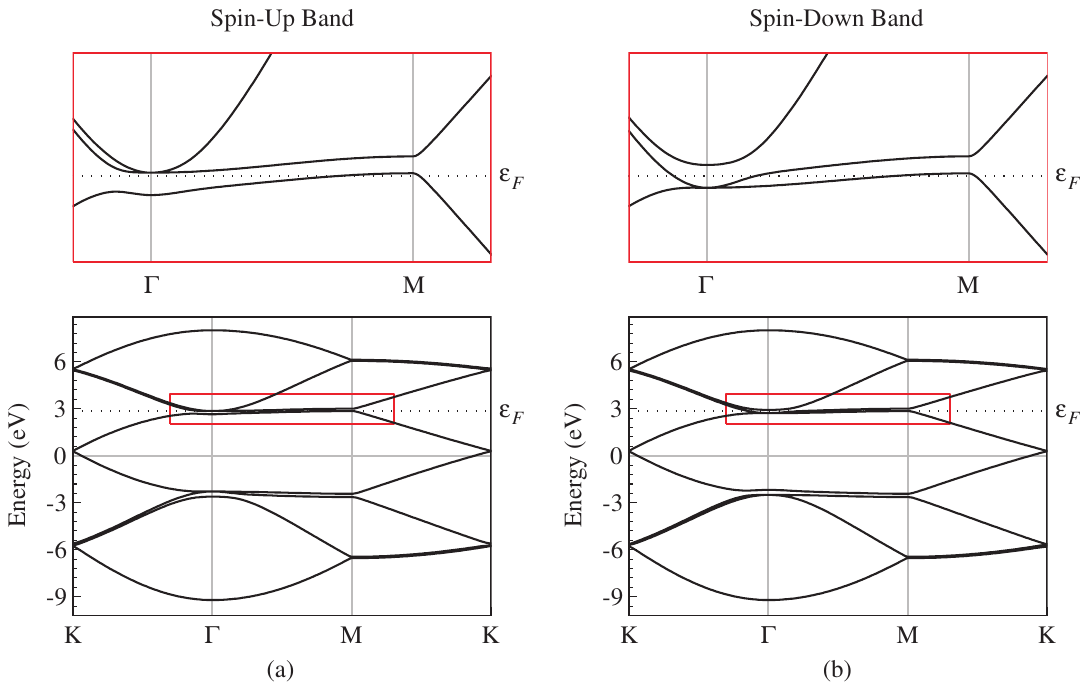}
	\end{center}
	\caption{Band structure of (a) spin-up and (b) spin-down electrons in the collinear SDW state at $\alpha  = 0.0,\delta  = 0.25$. The upper panels are zoom-in images of the regions surrounded by red rectangles in the lower panels.}
	\label{fig4}
\end{figure}

\begin{figure}[h!]
	\begin{center}
		\includegraphics[width=14.0cm]{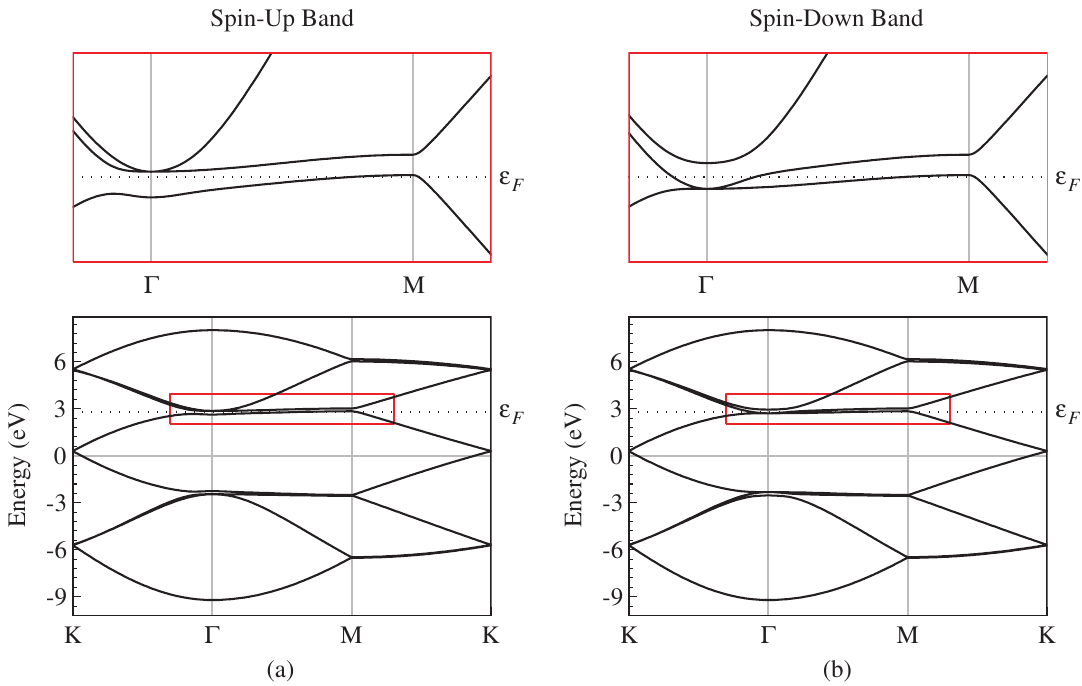}
	\end{center}
	\caption{Band structure of (a) spin-up and (b) spin-down electrons in the collinear SBO state at $\alpha  = 1.0,\delta  = 0.25$. The upper panels are zoom-in images of the regions surrounded by red rectangles in the lower panels.}
	\label{fig5}
\end{figure}

The band structure for the collinear SDW state at the parameter set of $\alpha  = 0.0,\delta  = 0.25$ is shown in Fig. \ref{fig4}. Due to small value of $\Delta$, bands formed by spin-up and spin-down electrons are almost identical. However, by looking at them carefully (see the upper panels of Fig. \ref{fig4}) and comparing the values in Table \ref{tab02}, we can find that the 5-th and the 6-th spin-down bands are overlapped, while a gap of 4.7meV exists between those spin-up bands. The band structure for the collinear SBO state at $\alpha  = 1.0,\delta  = 0.25$ is presented in Fig. \ref{fig5}. The 5-th and 6-th bands of spin-down electrons are overlapped, while a gap of 31.3meV exists between those bands of spin-up electrons (see Table \ref{tab02}).

\begin{figure}[h!]
	\begin{center}
		\includegraphics[width=14.0cm]{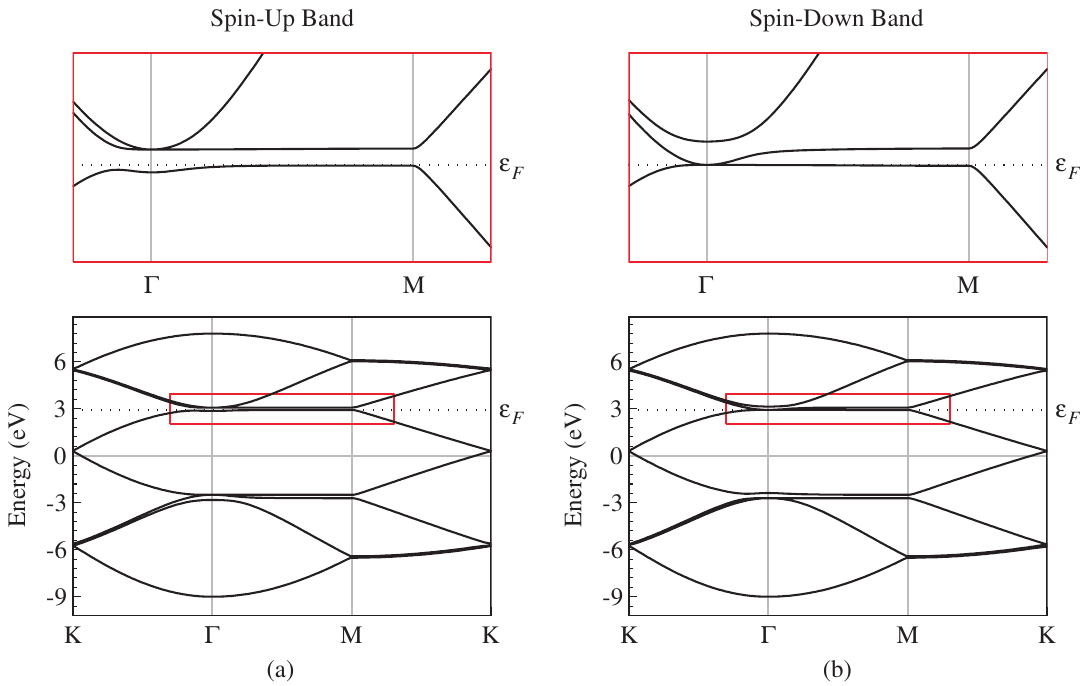}
	\end{center}
	\caption{Band structure of (a) spin-up and (b) spin-down electrons in the collinear SDW state at $\alpha  = 0.0,\delta  = 0.25$, with $t_3$ changed to zero. The upper panels are zoom-in images of the lower panels.}
	\label{fig6}
\end{figure}

\begin{figure}[h!]
	\begin{center}
		\includegraphics[width=14.0cm]{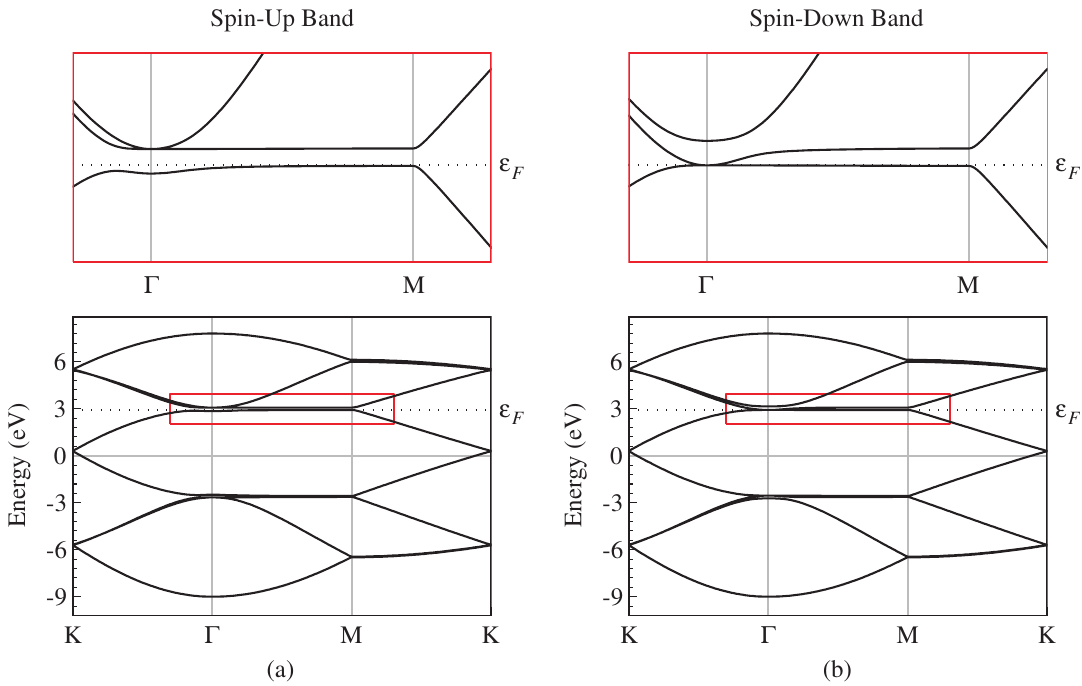}
	\end{center}
	\caption{Band structure of (a) spin-up and (b) spin-down electrons in the collinear SBO state at $\alpha  = 1.0,\delta  = 0.25$, with $t_3$ changed to zero. The upper panels are zoom-in images of the lower panels.}
	\label{fig7}
\end{figure}

The band structure of the \emph{half-metallic} nature has been found in the previous study on the collinear SDW\cite{ref18}. Namely, spin-down electrons have gapless excitations, but spin-up electrons are fully gapped, which leads to a metal of spin-down electrons and an insulator of spin-up electrons. In contrast with this, both electrons have gapless excitations in our result, as shown in Figs. \ref{fig4} and \ref{fig5}. More precisely, for both cases of the SDW and the SBO, spin-up electrons have the Fermi surface with three hole-pockets at the ${\bf{M}}$ points, while spin-down electrons have three hole-pockets at ${\bf{M}}$ and an electron-pocket at the ${\bf{\Gamma}}$ point. Thus, the half-metallic state was not found in this work.

We consider again two parameter sets, ($\alpha=0.0,\delta=0.25$) and ($\alpha=1.0,\delta=0.25$), with $t_3$ set to zero. As mentioned above, the chiral spin orders are favored in these cases. However, we have shown in Figs. \ref{fig6} and \ref{fig7} the band structures of the collinear SDW and SBO states, for a comparison with results in Ref. \cite{ref18}. As one can see from Fig. \ref{fig6}, the half-metallic collinear SDW, which has been proposed in the previous work\cite{ref18}, is reproduced here under the condition of $t_3=0$. This fact demonstrates clearly the effect of the hopping parameter $t_3$. In the absence of the parameter $t_3$, the Fermi surface of graphene in the original BZ is ideally nested, which manifests itself as a flat dispersion on the $\bf{\Gamma}-\bf{M}$ line of the small BZ. This flat part is favorable for gaping out the bands near the Fermi level $\epsilon_F$. As a result, the chiral spin-ordered states become insulators with the lower energies than the half-metallic collinear ones. Therefore, the chiral SDW and SBO states are realized when $t_3$ is eliminated. However, in the presence of $t_3$, the Fermi surface is away from the nesting and the band dispersion has a slope on the $\bf{\Gamma}-\bf{M}$ line, which prevents the formation of the gap near $\epsilon_F$. Thus, the collinear states become metallic by the effect of the third-nearest-neighbor hopping.

Finally, we present the real-space representation of $H_{{\rm{SPN}}}$, which is probably helpful for the understanding of the collinear SBO. Inserting $c_{{\bf{k}},o,\sigma }  = \frac{1}{{\sqrt N }}\sum\limits_i {c_{i,o,\sigma } e^{ - i{\bf{k}} \cdot {\bf{R}}_i } }$ into Eq. (\ref{eq22}), we have the real-space representation of $H_{{\rm{SPN}}}$,
\begin{eqnarray}\label{eq24}
\eqalign{
H_{{\rm{SPN}}}  = E_{{\rm{SPN}}}  - 2\sum\limits_{o,o',m} {\sum\limits_i {\vec \Delta _{o'o} ({\bf{R}}_i  - {\bf{R}}_m ,{\bf{R}}_i ) \cdot } } \left( {\sum\limits_{\sigma ,\sigma '} {c_{{\bf{R}}_i  - {\bf{R}}_m ,o',\sigma }^ \dagger  \vec \sigma _{\sigma \sigma '} c_{{\bf{R}}_i ,o,\sigma '} } } \right), \\
\vec \Delta _{o'o} ({\bf{R}}_i  - {\bf{R}}_m ,{\bf{R}}_i ) \equiv \sum\limits_{\alpha  = 1}^3 {e^{ - i{\bf{M}}_\alpha   \cdot {\bf{R}}_i } \left[ {\varphi _{oo'm}^{\rm{C}} ({\bf{M}}_\alpha  )} \right]^* } \vec \Delta ^{{\rm{SPN}}} ({\bf{M}}_\alpha  ), \\
E_{{\rm{SPN}}}  \equiv \frac{{4N}}{\Lambda }\sum\limits_{\alpha  = 1}^3 {|\vec \Delta ^{{\rm{SPN}}} ({\bf{M}}_\alpha  )|^2 }.	
}
\end{eqnarray}
Let us consider the spatial pattern of the order parameter $\vec \Delta _{o'o} ({\bf{R}}_i  - {\bf{R}}_m ,{\bf{R}}_i )$ for both spin orders, the collinear SDW and the collinear SBO.

\begin{figure}[h!]
	\begin{center}
		\includegraphics[width=8.6cm]{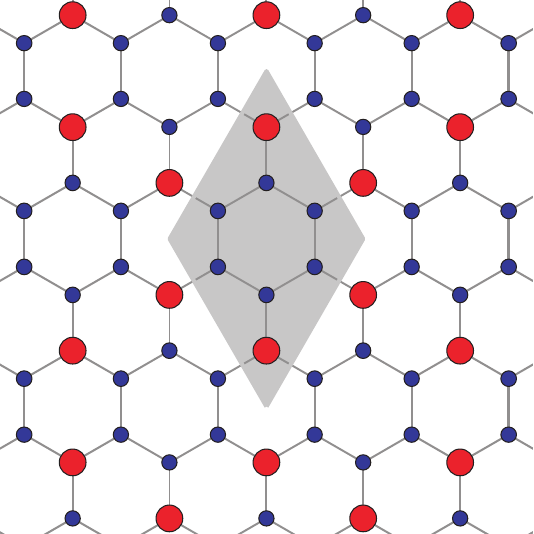}
	\end{center}
	\caption{Real-space pattern of the order parameter $\Delta _o ({\bf{R}}_i)$ for the collinear SDW at $\alpha=0.0,\delta=0.25$. The red (blue) circle represents the positive (negative) value of $\Delta _o ({\bf{R}}_i)$, while its radius is proportional to $\left| {\Delta _o ({\bf{R}}_i )} \right|$. The gray-shaded area indicates the enlarged unit cell of this spin-ordered state.}
	\label{fig8}
\end{figure}

In the case of the SDW, the irreducible singular modes in Fig. \ref{fig2} are approximated by
\begin{eqnarray}\label{eq25}
\eqalign{
\varphi _{AA1}^{\rm{C}} ({\bf{M}}_1 ) = \varphi _{BB1}^{\rm{C}} ({\bf{M}}_1 ) = \varphi _{AA1}^{\rm{C}} ({\bf{M}}_2 ) =  - \varphi _{BB1}^{\rm{C}} ({\bf{M}}_2 )\\
\hspace{2pc} = \varphi _{AA1}^{\rm{C}} ({\bf{M}}_3 ) =  - \varphi _{BB1}^{\rm{C}} ({\bf{M}}_3 ) = \phi,\\
\textrm{other elements} \approx 0,
}
\end{eqnarray}
with a positive number $\phi$ and the lattice vector ${\bf{R}}_1  = (0,0)$. The above equation gives the following result for the collinear SDW:
\begin{eqnarray}\label{eq26}
\eqalign{
\vec \Delta _{oo} ({\bf{R}}_i ,{\bf{R}}_i ) = \Delta _o ({\bf{R}}_i ){\bf{e}}_z ,\\
\Delta _A ({\bf{R}}_i ) = \phi \Delta (e^{ - i{\bf{M}}_1 \cdot {\bf{R}}_i } + e^{ - i{\bf{M}}_2 \cdot {\bf{R}}_i }+ e^{ - i{\bf{M}}_3  \cdot {\bf{R}}_i } ),\\
\Delta _B ({\bf{R}}_i ) = \phi \Delta (e^{ - i{\bf{M}}_1 \cdot {\bf{R}}_i } - e^{ - i{\bf{M}}_2 \cdot {\bf{R}}_i }- e^{ - i{\bf{M}}_3  \cdot {\bf{R}}_i } ), \textrm{others} \approx 0.
}
\end{eqnarray}
The real-space pattern of $\Delta _o ({\bf{R}}_i )$ is shown in Fig. \ref{fig8}.

\begin{figure}[h!]
	\begin{center}
		\includegraphics[width=8.6cm]{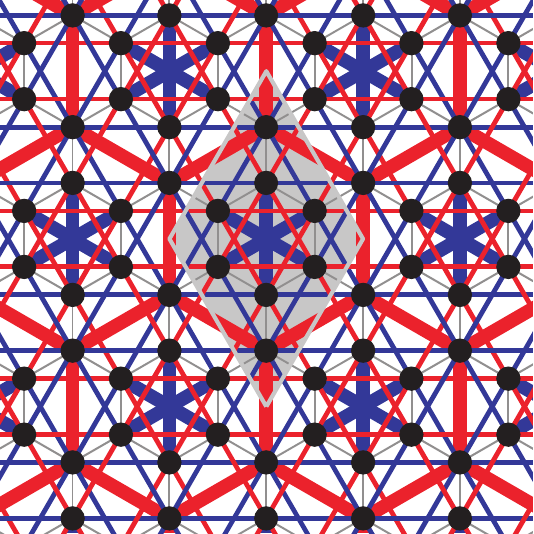}
	\end{center}
	\caption{Real-space pattern of the order parameter $\Delta _{o'o} ({\bf{R}}_j ,{\bf{R}}_i ) = \Delta _{oo'} ({\bf{R}}_i ,{\bf{R}}_j )$ for the collinear SBO at $\alpha=1.0,\delta=0.25$. The red (blue) stick represents the positive (negative) value of $\Delta _{oo'} ({\bf{R}}_i ,{\bf{R}}_j )$, while its width is proportional to $\left| {\Delta _{oo'} ({\bf{R}}_i ,{\bf{R}}_j )} \right|$. The gray-shaded area indicates the enlarged unit cell of this spin-ordered state.}
	\label{fig9}
\end{figure}

In the case of the SBO, the irreducible singular modes in Fig. \ref{fig3} are approximated by
\begin{eqnarray}\label{eq27}
\eqalign{
\varphi _{AB2}^{\rm{C}} ({\bf{M}}_1 ) = \varphi _{AB5}^{\rm{C}} ({\bf{M}}_1 ) = \varphi _{BA2}^{\rm{C}} ({\bf{M}}_1 ) = \varphi _{BA5}^{\rm{C}} ({\bf{M}}_1 )\\
\hspace{2pc} = \varphi _{AB2}^{\rm{C}} ({\bf{M}}_2 ) = \varphi _{AB9}^{\rm{C}} ({\bf{M}}_2 ) =  - \varphi _{BA5}^{\rm{C}} ({\bf{M}}_2 ) =  - \varphi _{BA12}^{\rm{C}} ({\bf{M}}_2 )\\
\hspace{2pc} = \varphi _{AB5}^{\rm{C}} ({\bf{M}}_3 ) = \varphi _{AB9}^{\rm{C}} ({\bf{M}}_3 ) = -\varphi _{BA2}^{\rm{C}} ({\bf{M}}_3 ) = - \varphi _{BA12}^{\rm{C}} ({\bf{M}}_3 )=\phi_1, \\
- \varphi _{AA2}^{\rm{C}} ({\bf{M}}_1 ) =  - \varphi _{AA5}^{\rm{C}} ({\bf{M}}_1 ) =  - \varphi _{BB2}^{\rm{C}} ({\bf{M}}_1 ) = - \varphi _{BB5}^{\rm{C}} ({\bf{M}}_1 )\\
\hspace{2pc} = - \varphi _{AA4}^{\rm{C}} ({\bf{M}}_2 ) = - \varphi _{AA7}^{\rm{C}} ({\bf{M}}_2 ) = \varphi _{BB4}^{\rm{C}} ({\bf{M}}_2 ) = \varphi _{BB7}^{\rm{C}} ({\bf{M}}_2 )\\
\hspace{2pc} = - \varphi _{AA3}^{\rm{C}} ({\bf{M}}_3 ) = -\varphi _{AA6}^{\rm{C}} ({\bf{M}}_3 ) = \varphi_{BB3}^{\rm{C}} ({\bf{M}}_3 ) = \varphi _{BB6}^{\rm{C}} ({\bf{M}}_3 ) = \phi_2,\\
\textrm{other elements} \approx 0.
}
\end{eqnarray}
Here $\phi_1$ and $\phi_2 ( < \phi_1 )$ are positive numbers, and the lattice vectors ${\bf{R}}_i$ are expressed as
\begin{eqnarray}\nonumber
\eqalign{
{\bf{R}}_2  =  - {\bf{R}}_5  = a(1,0),{\bf{R}}_3  =  - {\bf{R}}_6  = a(1/2,\sqrt 3 /2),\\
{\bf{R}}_4  =  - {\bf{R}}_7  = a( - 1/2,\sqrt 3 /2),{\bf{R}}_9  =  - {\bf{R}}_{12}  = a(0,\sqrt 3 ).
}
\end{eqnarray}
The following result for the collinear SBO is obtained from Eq. (\ref{eq27}).
\begin{eqnarray}\label{eq28}
\eqalign{
\vec \Delta _{o'o} ({\bf{R}}_i  - {\bf{R}}_m ,{\bf{R}}_i ) = \Delta _{o'o} ({\bf{R}}_i  - {\bf{R}}_m ,{\bf{R}}_i ){\bf{e}}_z ,\\
\Delta _{BA} ({\bf{R}}_i  - {\bf{R}}_2 ,{\bf{R}}_i ) = \Delta _{AB} ({\bf{R}}_i ,{\bf{R}}_i  - {\bf{R}}_2 ) = \phi _1 \Delta (e^{ - i{\bf{M}}_1  \cdot {\bf{R}}_i }  + e^{ - i{\bf{M}}_2  \cdot {\bf{R}}_i } ),\\
\Delta _{BA} ({\bf{R}}_i  - {\bf{R}}_5 ,{\bf{R}}_i ) = \Delta _{AB} ({\bf{R}}_i ,{\bf{R}}_i  - {\bf{R}}_5 ) = \phi _1 \Delta (e^{ - i{\bf{M}}_1  \cdot {\bf{R}}_i }  + e^{ - i{\bf{M}}_3  \cdot {\bf{R}}_i } ),\\
\Delta _{BA} ({\bf{R}}_i  - {\bf{R}}_9 ,{\bf{R}}_i ) = \Delta _{AB} ({\bf{R}}_i ,{\bf{R}}_i  - {\bf{R}}_9 ) = \phi _1 \Delta (e^{ - i{\bf{M}}_2  \cdot {\bf{R}}_i }  + e^{ - i{\bf{M}}_3  \cdot {\bf{R}}_i } ),\\
\Delta _{AA} ({\bf{R}}_i  - {\bf{R}}_2 ,{\bf{R}}_i ) = \Delta _{AA} ({\bf{R}}_i ,{\bf{R}}_i  - {\bf{R}}_2 ) =  - \phi _2 \Delta e^{ - i{\bf{M}}_1  \cdot {\bf{R}}_i } ,\\
\Delta _{BB} ({\bf{R}}_i  - {\bf{R}}_2 ,{\bf{R}}_i ) = \Delta _{BB} ({\bf{R}}_i ,{\bf{R}}_i  - {\bf{R}}_2 ) =  - \phi _2 \Delta e^{ - i{\bf{M}}_1  \cdot {\bf{R}}_i } ,\\
\Delta _{AA} ({\bf{R}}_i  - {\bf{R}}_3 ,{\bf{R}}_i ) = \Delta _{AA} ({\bf{R}}_i ,{\bf{R}}_i  - {\bf{R}}_3 ) =  - \phi _2 \Delta e^{ - i{\bf{M}}_3  \cdot {\bf{R}}_i } ,\\
\Delta _{BB} ({\bf{R}}_i  - {\bf{R}}_3 ,{\bf{R}}_i ) = \Delta _{BB} ({\bf{R}}_i ,{\bf{R}}_i  - {\bf{R}}_3 ) = \phi _2 \Delta e^{ - i{\bf{M}}_3  \cdot {\bf{R}}_i } ,\\
\Delta _{AA} ({\bf{R}}_i  - {\bf{R}}_4 ,{\bf{R}}_i ) = \Delta _{AA} ({\bf{R}}_i ,{\bf{R}}_i  - {\bf{R}}_4 ) =  - \phi _2 \Delta e^{ - i{\bf{M}}_2  \cdot {\bf{R}}_i } ,\\
\Delta _{BB} ({\bf{R}}_i  - {\bf{R}}_4 ,{\bf{R}}_i ) = \Delta _{BB} ({\bf{R}}_i ,{\bf{R}}_i  - {\bf{R}}_4 ) = \phi _2 \Delta e^{ - i{\bf{M}}_2  \cdot {\bf{R}}_i } , \textrm{others} \approx 0.
}
\end{eqnarray}
The real-space pattern of $\Delta _{o'o} ({\bf{R}}_j ,{\bf{R}}_i ) = \Delta _{oo'} ({\bf{R}}_i ,{\bf{R}}_j )$ is shown in Fig. \ref{fig9}.

It is a remarkable feature of the collinear SBO state that the strongest spin bonds are formed between the third-nearest neighbors, not between the nearest neighbors. This abnormal property has not been reported before. In many previous SBO states, the strongest bonds are formed between the nearest neighbors. For example, the SBO state of the half-filled honeycomb lattice, dubbed spin-Kekul\'{e} phase\cite{ref21, ref31}, possesses the strongest spin bonds between the nearest neighbors. The same is true for the SBO phase of the kagome lattice \cite{ref40, ref45, ref46}.

\section{Conclusion}\label{sec5}

In this paper, we considered graphene doped close to the VHS. Many-body instabilities of correlated electrons were identified by using TUFRG, and the results were summarized by the ground-state phase diagram in the space of the doping level and the screening parameter. With the interaction parameters suitable for graphene, we found the anomalous SBO phase at quarter doping, while the chiral $d$-wave SC around it. The SBO is transformed into the SDW phase when involving a weak additional screening. In the case of strong screening, the SDW and the chiral $d$-wave SC become main ingredients of the phase diagram.

We focused on the SDW and SBO states at quarter doping and used recently developed TUFRG + MF approach to describe in detail both of these spin-ordered states. The collinear SDW and SBO turn out to be energetically more stable compared to the chiral ones. But if the third-nearest-neighbor hopping is absent, the spin orders become chiral for both cases. The chiral and collinear orders are easily transformed into each other, depending on the shape of the Fermi surface.

Band structures of the collinear SDW and SBO states were presented. The 5-th and 6-th bands share the Fermi surface. In both cases, these two bands are overlapped for the spin projection along $\vec \Delta ^{{\rm{SPN}}} ({\bf{M}}_i )$, while gapped (4.7meV in the SDW and 31.3meV in the SBO) for the one opposite to $\vec \Delta ^{{\rm{SPN}}} ({\bf{M}}_i )$. Unlike the previous study\cite{ref18} where the half metal has been suggested, spin-up and spin-down bands have gapless excitations in our result. For both cases of the SDW and the SBO, spin-up band has the hole-pockets at ${\bf{M}}$, while spin-down band has the hole-pockets at ${\bf{M}}$ and the electron-pocket at ${\bf{\Gamma}}$. If the hopping parameter $t_3$ is removed, the half-metallic state will be recovered.

\ack{
We thank Kwang-Il Ryom and Myong-Chol Pak for useful discussions.
}

\section*{Data availability statement}
All data that support the findings of this study are included within the article.

\section*{References}
\bibliography{SJO_JPCM_2025V2}

\begin{thebibliography}{10}

\bibitem{ref01}
K.~S. Novoselov, A.~K. Geim, S.~V. Morozov, D.~Jiang, Y.~Zhang, S.~V. Dubonos,
  I.~V. Grigorieva, and A.~A. Firsov.
\newblock {\em Science}, 306:666, 2004.
\url{https://doi.org/10.1126/science.1102896}.

\bibitem{ref02}
A.~H.~Castro Neto, F.~Guinea, N.~M.~R. Peres, K.~S. Novoselov, and A.~K. Geim.
\newblock {\em Rev. Mod. Phys.}, 81:109, 2009.
\url{https://doi.org/10.1103/RevModPhys.81.109}.

\bibitem{ref02a}
W.~Zhao, S.~Zhao, H.~Li, S.~Wang, S.~Wang, M.~Utama, S.~Kahn, Y.~Jiang,
  X.~Xiao, S.~Yoo, K.~Watanabe, T.~Taniguchi, A.~Zettl, and F.~Wang.
\newblock {\em Nature}, 594:517, 2021.
\url{https://doi.org/10.1038/s41586-021-03574-4}.

\bibitem{ref02b}
S.~Fujiyama, H.~Maebashi, N.~Tajima, T.~Tsumuraya, H-B. Cui, M.~Ogata, and
  R.~Kato.
\newblock {\em Phys. Rev. Lett.}, 128:027201, 2022.
\url{https://doi.org/10.1103/PhysRevLett.128.027201}.

\bibitem{ref02e}
L.~Li, J.~G. Checkelsky, Y.~S. Hor, C.~Uher, A.~F. Hebard, R.~J. Cava, and
  N.~P. Ong.
\newblock {\em Science}, 321:547, 2008.
\url{https://doi.org/10.1126/science.1158908}.

\bibitem{ref02c}
P.~Li, B.~Liu, S.~Chen, W.-X. Zhang, and Z.-X. Guo.
\newblock {\em Chin. Phys. B}, 33:017505, 2024.
\url{https://doi.org/10.1088/1674-1056/acf65f}.

\bibitem{ref02d}
P.~Li, X.~Yang, Q.-S. Jiang, Y.-Z. Wu, and W.~Xun.
\newblock {\em Phys. Rev. Mater.}, 7:064002, 2023.
\url{https://doi.org/10.1103/PhysRevMaterials.7.064002}.

\bibitem{ref03}
O.~Vafek.
\newblock {\em Phys. Rev. Lett.}, 98:216401, 2007.
\url{https://doi.org/10.1103/PhysRevLett.98.216401}.

\bibitem{ref04}
D.~E. Sheehy and J.~Schmalian.
\newblock {\em Phys. Rev. Lett.}, 99:226803, 2007.
\url{https://doi.org/10.1103/PhysRevLett.99.226803}.

\bibitem{ref05}
P.-S. He, S.-J. Oh, Y.~Chen, and G.-S. Tian.
\newblock {\em Commun. Theor. Phys.}, 54:897, 2010.
\url{https://doi.org/10.1088/0253-6102/54/5/24}.

\bibitem{ref06}
X.~Zhou, L.-J. Wan, and Y.-G. Guo.
\newblock {\em Adv. Mater.}, 25:2152, 2013.
\url{https://doi.org/10.1002/adma.201300071}.

\bibitem{ref07}
N.~M. Gabor, J.~C.~W. Song, Q.~Ma, N.~L. Nair, T.~Taychatanapat, K.~Watanabe,
  T.~Taniguchi, L.~S. Levitov, and P.~Jarillo-Herrero.
\newblock {\em Science}, 334:648, 2011.
\url{https://doi.org/10.1126/science.1211384}.

\bibitem{ref08}
N.~Furukawa, T.~M. Rice, and M.~Salmhofer.
\newblock {\em Phys. Rev. Lett.}, 81:3195, 1998.
\url{https://doi.org/10.1103/PhysRevLett.81.3195}.

\bibitem{ref09}
A.~M. Black-Schaffer and C.~Honerkamp.
\newblock {\em J. Phys.: Condens. Matter}, 26:423201, 2014.
\url{https://doi.org/10.1088/0953-8984/26/42/423201}.

\bibitem{ref10}
J.~Gonz\'{a}lez.
\newblock {\em Phys. Rev. B}, 78:205431, 2008.
\url{https://doi.org/10.1103/PhysRevB.78.205431}.

\bibitem{ref11}
Z.-C. Gu, H.-C. Jiang, D.~N. Sheng, H.~Yao, L.~Balents, and X.-G. Wen.
\newblock {\em Phys. Rev. B}, 88:155112, 2013.
\url{https://doi.org/10.1103/PhysRevB.88.155112}.

\bibitem{ref12}
T.~Ying and S.~Wessel.
\newblock {\em Phys. Rev. B}, 97:075127, 2018.
\url{https://doi.org/10.1103/PhysRevB.97.075127}.

\bibitem{ref13}
P.~Jia, S.~Yang, W.~Li, J.~Yang, T.~Ying, X.~Li, and X.~Sun.
\newblock {\em Phys. Lett. A}, 442:128175, 2022.
\url{https://doi.org/10.1016/j.physleta.2022.128175}.

\bibitem{ref14}
R.~Nandkishore, L.~S. Levitov, and A.~V. Chubukov.
\newblock {\em Nat. Phys.}, 8:158, 2012.
\url{https://doi.org/10.1038/nphys2208}.

\bibitem{ref15}
M.~L. Kiesel, C.~Platt, W.~Hanke, D.~A. Abanin, and R.~Thomale.
\newblock {\em Phys. Rev. B}, 86:020507(R), 2012.
\url{https://doi.org/10.1103/PhysRevB.86.020507}.

\bibitem{ref16}
W.-S. Wang, Y.-Y. Xiang, Q.-H. Wang, F.~Wang, F.~Yang, and D.-H. Lee.
\newblock {\em Phys. Rev. B}, 85:035414, 2012.
\url{https://doi.org/10.1103/PhysRevB.85.035414}.

\bibitem{ref17}
T.~Li.
\newblock {\em Europhys. Lett.}, 97:37001, 2012.
\url{https://doi.org/10.1209/0295-5075/97/37001}.

\bibitem{ref18}
R.~Nandkishore, G.~W. Chern, and A.~V. Chubukov.
\newblock {\em Phys. Rev. Lett.}, 108:227204, 2012.
\url{https://doi.org/10.1103/PhysRevLett.108.227204}.

\bibitem{ref19}
S.-J. O.
\newblock {\em Phys. Rev. B}, 109:205118, 2024.
\url{https://doi.org/10.1103/PhysRevB.109.205118}.

\bibitem{ref20}
J.~Lichtenstein, D.~S. de~la Pe\~{n}a, D.~Rohe, E.~D. Napoli, C.~Honerkamp, and
  S.~A. Maier.
\newblock {\em Comput. Phys. Commun.}, 213:100, 2017.
\url{https://doi.org/10.1016/j.cpc.2016.12.013}.

\bibitem{ref21}
Y.-U. An, S.-J. O, K.-I. Ryom, and I.-G. Son.
\newblock {\em Physica B (Amsterdam)}, 655:414748, 2023.
\url{https://doi.org/10.1016/j.physb.2023.414748}.

\bibitem{ref22}
T.~O. Wehling, E.~\c{S}a\c{s}\i{}o\u{g}lu, C.~Friedrich, A.~I. Lichtenstein,
  M.~I. Katsnelson, and S.~Bl\"{u}gel.
\newblock {\em Phys. Rev. Lett.}, 106:236805, 2011.
\url{https://doi.org/10.1103/PhysRevLett.106.236805}.

\bibitem{refA1}
P.~Rosenzweig, H.~Karakachian, D.~Marchenko, K.~K\"{u}ster, and U.~Starke.
\newblock {\em Phys. Rev. Lett.}, 125:176403, 2020.
\url{https://doi.org/10.1103/PhysRevLett.125.176403}.

\bibitem{refA2}
S.~A. Herrera, G.~Parra-Martinez, P.~Rosenzweig, B.~Matta, C.~M. Polley,
  K.~K\"{u}ster, U.~Starke, F.~Guinea, J.~A. Silva-Guillen, G.~G. Naumis, and
  P.~A. Pantaleon.
\newblock {\em ACS Nano}, 18:34842, 2024.
\url{https://doi.org/10.1021/acsnano.4c12532}.

\bibitem{ref23}
D.~K. Efetov and P.~Kim.
\newblock {\em Phys. Rev. Lett.}, 105:256805, 2010.
\url{https://doi.org/10.1103/PhysRevLett.105.256805}.

\bibitem{ref24}
W.~Metzner, M.~Salmhofer, C.~Honerkamp, V.~Meden, and K.~Sch\"{o}nhammer.
\newblock {\em Rev. Mod. Phys.}, 84:299, 2012.
\url{https://doi.org/10.1103/RevModPhys.84.299}.

\bibitem{ref25}
C.~Platt, W.~Hanke, and R.~Thomale.
\newblock {\em Adv. Phys.}, 62:453, 2013.
\url{https://doi.org/10.1080/00018732.2013.862020}.

\bibitem{ref26}
N.~Dupuis, L.~Canet, A.~Eichhorn, W.~Metzner, J.~M. Pawlowski, M.~Tissier, and
  N.~Wschebor.
\newblock {\em Phys. Rep.}, 910:1, 2021.
\url{https://doi.org/10.1016/j.physrep.2021.01.001}.

\bibitem{ref27}
C.~Husemann and M.~Salmhofer.
\newblock {\em Phys. Rev. B}, 79:195125, 2009.
\url{https://doi.org/10.1103/PhysRevB.79.195125}.

\bibitem{ref28}
D.~S. de~la Pe\~{n}a, J.~Lichtenstein, and C.~Honerkamp.
\newblock {\em Phys. Rev. B}, 95:085143, 2017.
\url{https://doi.org/10.1103/PhysRevB.95.085143}.

\bibitem{ref29}
D.~S. de~la Pe\~{n}a, J.~Lichtenstein, C.~Honerkamp, and M.~M. Scherer.
\newblock {\em Phys. Rev. B}, 96:205155, 2017.
\url{https://doi.org/10.1103/PhysRevB.96.205155}.

\bibitem{ref30}
G.~A.~H. Schober, J.~Ehrlich, T.~Reckling, and C.~Honerkamp.
\newblock {\em Front. Phys.}, 6:32, 2018.
\url{https://doi.org/10.3389/fphy.2018.00032}.

\bibitem{ref31}
S.-J. O, Y.-H. Kim, H.-Y. Rim, H.-C. Pak, and S.-J. Im.
\newblock {\em Phys. Rev. B}, 99:245140, 2019.
\url{https://doi.org/10.1103/PhysRevB.99.245140}.

\bibitem{ref32}
S.-J. O, Y.-H. Kim, O.-G. Pak, K.-H. Jong, C.-W. Ri, and H.-C. Pak.
\newblock {\em Phys. Rev. B}, 103:235150, 2021.
\url{https://doi.org/10.1103/PhysRevB.103.235150}.

\bibitem{ref33}
J.~Ehrlich and C.~Honerkamp.
\newblock {\em Phys. Rev. B}, 102:195108, 2020.
\url{https://doi.org/10.1103/PhysRevB.102.195108}.

\bibitem{ref34}
J.~B. Profe, C.~Honerkamp, S.~Achilles, and D.~M. Kennes.
\newblock {\em Phys. Rev. Res.}, 3:023180, 2021.
\url{https://doi.org/10.1103/PhysRevResearch.3.023180}.

\bibitem{ref35}
N.~Gneist, L.~Classen, and M.~M. Scherer.
\newblock {\em Phys. Rev. B}, 106:125141, 2022.
\url{https://doi.org/10.1103/PhysRevB.106.125141}.

\bibitem{ref36}
J.~Beyer, J.~B. Profe, L.~Klebl, T.~Schwemmer, D.~M. Kennes, R.~Thomale,
  C.~Honerkamp, and S.~Rachel.
\newblock {\em Phys. Rev. B}, 107:125115, 2023.
\url{https://doi.org/10.1103/PhysRevB.107.125115}.

\bibitem{ref37}
N.~K. Yirga, K.-M. Tam, and D.~K. Campbell.
\newblock {\em Phys. Rev. B}, 107:235120, 2023.
\url{https://doi.org/10.1103/PhysRevB.107.235120}.

\bibitem{ref38}
J.~B. Profe, S.~Beck, D.~M. Kennes, A.~Georges, and O.~Gingras.
\newblock {\em npj Quantum Mater.}, 9:53, 2024.
\url{https://doi.org/10.1038/s41535-024-00661-3}.

\bibitem{ref39}
P.~M. Bonetti, D.~Chakraborty, X.~Wu, and A.~P. Schnyder.
\newblock {\em Phys. Rev. B}, 109:L180509, 2024.
\url{https://doi.org/10.1103/PhysRevB.109.L180509}.

\bibitem{ref40}
J.~B. Profe, L.~Klebl, F.~Grandi, H.~Hohmann, M.~D\"{u}rrnagel, T.~Schwemmer,
  R.~Thomale, and D.~M. Kennes.
\newblock {\em Phys. Rev. Res.}, 6:043078, 2024.
\url{https://doi.org/10.1103/PhysRevResearch.6.043078}.

\bibitem{ref41}
J.~B. Profe and D.~M. Kennes.
\newblock {\em Eur. Phys. J. B}, 95:60, 2022.
\url{https://doi.org/10.1140/epjb/s10051-022-00316-x}.

\bibitem{ref42}
J.~Beyer, J.~B. Profe, and L.~Klebl.
\newblock {\em Eur. Phys. J. B}, 95:65, 2022.
\url{https://doi.org/10.1140/epjb/s10051-022-00323-y}.

\bibitem{ref43}
N.~Gneist, D.~Kiese, R.~Henkel, R.~Thomale, L.~Classen, and M.~M. Scherer.
\newblock {\em Eur. Phys. J. B}, 95:157, 2022.
\url{https://doi.org/10.1140/epjb/s10051-022-00395-w}.

\bibitem{ref44}
J.~Wang, A.~Eberlein, and W.~Metzner.
\newblock {\em Phys. Rev. B}, 89:121116(R), 2014.
\url{https://doi.org/10.1103/PhysRevB.89.121116}.

\bibitem{ref45}
W.-S. Wang, Z.-Z. Li, Y.-Y. Xiang, and Q.-H. Wang.
\newblock {\em Phys. Rev. B}, 87:115135, 2013.
\url{https://doi.org/10.1103/PhysRevB.87.115135}.

\bibitem{ref46}
M.~L. Kiesel, C.~Platt, and R.~Thomale.
\newblock {\em Phys. Rev. Lett.}, 110:126405, 2013.
\url{https://doi.org/10.1103/PhysRevLett.110.126405}.

\end{thebibliography}

\end{document}